\documentclass[twocolumn,showpacs,preprintnumbers,amsmath,amssymb,prl]{revtex4}
\usepackage{graphicx}
\usepackage{dcolumn}
\usepackage{bm}

\begin{document}

\title{Nonlinear microwave response of MgB$_2$}

\author{T.~Dahm}
\email{thomas.dahm@uni-tuebingen.de}
\affiliation{Institut f\"ur Theoretische Physik, 
         Universit\"at T\"ubingen, 
         Auf der Morgenstelle 14, D-72076 T\"ubingen, 
         Germany}

\author{D.~J.~Scalapino}
\email{djs@vulcan2.physics.ucsb.edu}
\affiliation{Department of Physics, University of California, Santa
  Barbara, California 93106-9530, U.S.A.} 

\date{\today}

\begin{abstract}
We calculate the intrinsic nonlinear microwave response of the two
gap superconductor MgB$_2$ in the clean and dirty limits. Due to the
small value of the $\pi$ band gap, the nonlinear response at low temperatures
is larger than for a single gap Bardeen-Cooper-Schrieffer (BCS) 
$s$-wave superconductor with a transition temperature of 40~K. Comparing
this result with the intrinsic nonlinear $d$-wave response of YBa$_2$Cu$_3$O$_7$
(YBCO) we find a comparable response at temperatures around 20~K. Due to its
two gap nature, impurity scattering in MgB$_2$ can be used to reduce the
nonlinear response if the scattering rate in the $\pi$ band
is made larger than the one in the $\sigma$ band.
\end{abstract}


\maketitle

High-$T_c$ cuprate thin films are being used for constructing
microstrip resonators and filters in the microwave regime. One of the limiting
factors is their nonlinear response leading to undesirable harmonic
generation and intermodulation. While this nonlinear response is often
associated with weak links, there is an intrinsic nonlinearity which sets a
lower limit on what can be achieved. This has previously been discussed for
the cuprates where the d-wave nature of the gap leads to an increase in the
intrinsic nonlinear response at low temperatures \cite{XuYipSauls,Dahm2}. 
The recent discovery of 
$s$-wave superconductivity 
in MgB$_2$ with a comparatively high critical temperature of $T_c=40$~K 
and progress in thin film preparation has led to the possibility of making
high $Q$ MgB$_2$ microstrip structures operating at 20 to 30 K. This raises
the question of how the nonlinearity of MgB$_2$ compares with that of the
cuprate superconductors \cite{Lamura,Booth}. 

Here, we study the intrinsic nonlinear microwave response of MgB$_2$.
By now it is well established that MgB$_2$ is a superconductor with two
different superconducting gaps associated with different parts of the
Fermi surface: a small gap ($\sim$ 2~meV) on the $\pi$ band and a 
large gap ($\sim$ 7~meV) on the $\sigma$ band \cite{Liu,Choi}. In a superconductor
the intrinsic nonlinear response arises from the backflow of
excited quasiparticles at finite temperatures.
The total current density $\vec{j}$ can be written as
$ \vec{j} = n e \vec{v}_s - \vec{j}_{\mathrm qp}$
where $\vec{v}_s$ is the superfluid velocity and $\vec{j}_{\mathrm qp}$
the quasiparticle backflow.
For a two band system like MgB$_2$ the quasiparticle backflow
consists of two contributions, one from each band with
$
\vec{j}_{\mathrm qp}=\vec{j}_{\mathrm qp,\sigma} + \vec{j}_{\mathrm qp,\pi} 
$.
In the clean limit these are given by
\begin{eqnarray}
 \vec{j}_{\mathrm qp,\alpha} &=& - 2 e N_\alpha(0) \int_{-\infty}^\infty
d\epsilon \Big\langle \vec{v}_{F,\alpha} f \left( \sqrt{\epsilon^2+
\Delta_\alpha^2\left( T \right)} \right. \nonumber \\
&& \left. + m \vec{v}_{F,\alpha} \cdot \vec{v}_s
\right) \Big\rangle_{FS,\alpha}
\label{qpback2b}
\end{eqnarray}
Here, $\alpha={\pi,\sigma}$ denotes the two bands in MgB$_2$, 
$\vec{v}_{F,\alpha}$ the Fermi velocity in each band, 
$\Delta_\alpha \left( T \right)$ the gaps, $f$ the Fermi function, 
$N_\alpha(0)$ the partial densities of states, and
$\langle \cdots \rangle_{FS,\alpha}$ a Fermi surface average over
band $\alpha$. The term $m \vec{v}_{F,\alpha} \cdot \vec{v}_s$
describes the Doppler shift in energy of the quasiparticles with 
respect to the superflow.

In MgB$_2$ the Fermi surfaces of the two bands have significantly
different topologies. In order to perform the Fermi surface averages
in \eqref{qpback2b} we use the band structure based model from
\cite{Dahm1}, in which the $\sigma$ band is modeled as a 
distorted cylinder and the $\pi$ band as a half torus.
For the temperature dependence of the two gaps we are using the
parameterization of Choi {\it et.~al} \cite{Choi} 
$\Delta_\sigma(T) = \Delta_\sigma(0) \sqrt{1-\left( T/T_c \right)^{2.9} }
$ and
$\Delta_\pi(T) = \Delta_\pi(0) \sqrt{1-\left( T/T_c \right)^{1.8} }$
with $\Delta_\sigma(0)=$6.8~meV and $\Delta_\pi(0)=$1.8~meV. This was found
to provide a reasonable fit to the anisotropic Eliashberg calculations of  
\cite{Choi}.

\begin{figure}
  \begin{center}
    \includegraphics[width=0.65\columnwidth,angle=270]{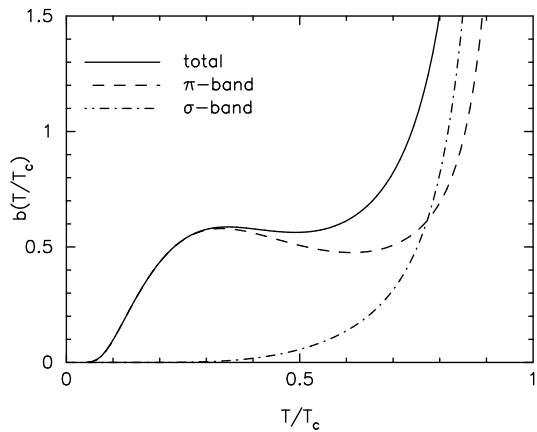}
    \caption{Temperature dependence of the total nonlinear coefficient $b(T)$
     for MgB$_2$ (solid line). The two contributions from the $\pi$ band
	and the $\sigma$ band are shown as the dashed and dashed-dotted line,
	respectively.
     \label{fig1} }
  \end{center}
\end{figure}

For currents flowing in the Boron plane, we 
expand Eq.~\eqref{qpback2b} up to third order in the current density.
This gives the leading $j^2$ nonlinear change of the superfluid density 
\cite{Dahm2,Dahm3}
\begin{equation}
 n_s(T,j) = n_{s0}(T) \left[ 1
- \frac{j^2}{j_{c,\pi}^2} \left( b_\pi(T) + 
\frac{j_{c,\pi}^2}{j_{c,\sigma}^2} b_\sigma(T) \right) \right]
\end{equation}
Here, the pair-breaking current densities of the two bands are
given by $j_{c,\alpha} = e N_\alpha(0) v_{F,\alpha} \Delta_\alpha(0)$.
As discussed in \cite{Dahm2,Dahm3} the nonlinear coefficients $b_\alpha(T)$ 
determine the strength of the nonlinear response as a function
of temperature and are given by
\begin{eqnarray}
b_\alpha(T) &=& - \frac{g_\alpha}{4} \Delta^2_\alpha(0) 
\left( \frac{n_\alpha}{n_{s0}(T)} \right)^3 \\
&& \int_{0}^\infty d\epsilon  \frac{d^3 f}{d E^3} \left( \sqrt{\epsilon^2+
\Delta_\alpha^2\left( T \right)} \right) \nonumber
\end{eqnarray}
Here, $g_\alpha$ is a geometrical Fermi surface factor with
$g_\pi$=3.36 and $g_\sigma$=1.
The linear part of the superfluid density $n_{s0}(T)$ is given by
\begin{equation}
 \frac{n_{s0}(T)}{n} = \sum_{\alpha=\pi,\sigma} \frac{n_\alpha}{n} \left( 1 +
2 \int_{0}^\infty d\epsilon  \frac{d f}{d E} \left( \sqrt{\epsilon^2+
\Delta_\alpha^2\left( T \right)} \right) \right) 
\end{equation}
where $n_\alpha$ is the electron density of the $\alpha$ band.
Using the known Fermi velocities and densities of states in MgB$_2$,
we find for the relative electron densities $n_\pi/n=0.692$ and
$n_\sigma/n=0.308$ and the pair-breaking current densities
$j_{c,\pi} = 3.32 \cdot 10^{8}  {\rm A/cm^2}$ and
$j_{c,\sigma} = 4.87 \cdot 10^{8}  {\rm A/cm^2}$.
Interestingly, these values turn out to be close to the estimate of the
pair-breaking current density of about $3 \cdot 10^{8}  {\rm A/cm^2}$ 
in YBCO, consistent with recent experiments \cite{Booth}. 
In Fig.~\ref{fig1} we show the temperature dependence of
the total nonlinear coefficient $b(T)=b_\pi(T) + 
\frac{j_{c,\pi}^2}{j_{c,\sigma}^2} b_\sigma(T)$ (solid line) along
with the contributions from the $\pi$ band (dashed) and the $\sigma$ band
(dashed-dotted). Due to its smaller gap, the $\pi$ band contribution
dominates at low temperatures and leads to  structure around 0.3$T_c$.
It is only at temperatures
above about 0.75$T_c$ that the contribution from the $\sigma$ band
becomes significant. The total nonlinear coefficient 
possesses a plateau-like region between 0.3 and 0.7$T_c$ in which
a reduction of the temperature does not improve the nonlinear response.

\begin{figure}
  \begin{center}
    \includegraphics[width=0.65\columnwidth,angle=270]{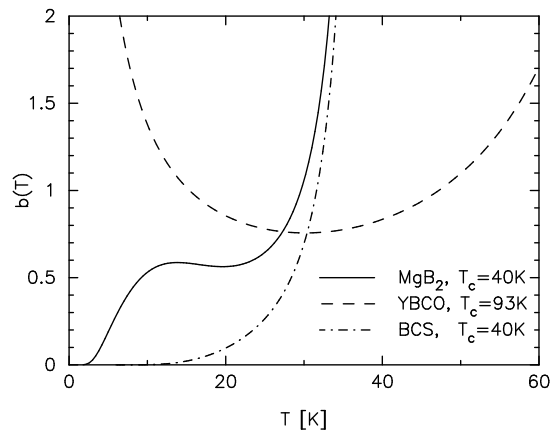}
    \caption{Temperature dependence of the intrinsic 
	nonlinear coefficient $b(T)$
	for MgB$_2$, for a $d$-wave superconductor with $T_c=93$~K
        (YBCO, dashed line), and a hypothetical BCS single gap superconductor 
	with $T_c=40$~K (dashed-dotted line).
     \label{fig2} }
  \end{center}
\end{figure}

In Fig.~\ref{fig2} we compare the nonlinear coefficient for MgB$_2$ 
(solid line) with
the intrinsic response for a $d$-wave superconductor with $T_c=93$~K
(YBCO, dashed line) and a hypothetical BCS single gap superconductor 
with $T_c=40$~K (dashed-dotted line)
assuming for simplicity that the pair-breaking current densities are the
same. For a $d$-wave superconductor $b(T)$ increases at low temperature
because of the gap nodes \cite{XuYipSauls,Dahm2}. For this reason the nonlinear
coefficient in MgB$_2$ becomes smaller than the one for YBCO at about
27~K. However, due to the presence of the small gap the nonlinear
response in MgB$_2$ is not as small as one would have expected for a
single gap $s$-wave superconductor (dashed-dotted line).

\begin{figure}
  \begin{center}
    \includegraphics[width=0.65\columnwidth,angle=270]{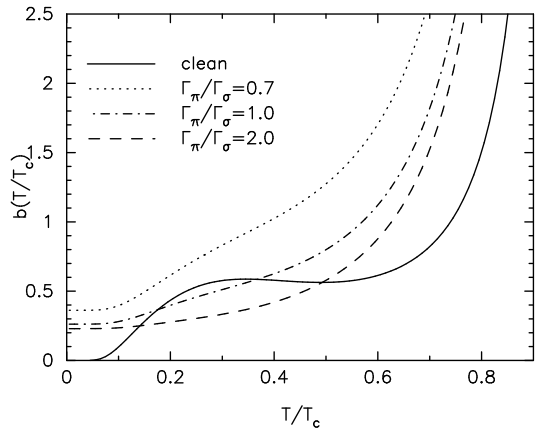}
    \caption{Temperature dependence of $b(T)$ in the clean (solid line)
     and dirty limit. In the dirty limit $b(T)$ depends on the relative
     scattering rates in the two bands of MgB$_2$. Results are shown for
     $\Gamma_\pi/\Gamma_\sigma=$0.7 (dotted), 1 (dashed-dotted), and 2 (dashed).
     \label{fig3} }
  \end{center}
\end{figure}

The foregoing analysis has been made in the clean limit without any
impurity scattering. However, in most current MgB$_2$ films, the 
scattering rate as judged from the residual
resistivity is larger than the two gaps \cite{JinAPL}. For this reason,
in the following we also want to discuss the dirty limit. In a two band
superconductor there are in principle three different scattering rates:
the two intraband scattering rates $\Gamma_\pi$ and $\Gamma_\sigma$ and
an interband scattering rate $\Gamma_{\pi\sigma}$. It has been shown
that because of the different parity of the local orbitals making up the
$\pi$ and $\sigma$ bands, the interband scattering rate is
much smaller than the intraband scattering rates \cite{Mazin}. 
Thus, in the following we will neglect $\Gamma_{\pi\sigma}$ and keep only
the intraband scattering rates. In the dirty limit the current
densities $\vec{j}_\alpha$ in the two bands are then given by \cite{XuYipSauls}
\begin{eqnarray}
 \vec{j}_{\alpha} &=& - i e N_\alpha(0) \pi T \sum_{n=-\infty}^\infty 
\nonumber\\
&& \left\langle \vec{v}_{F,\alpha} \frac{\tilde{\omega}_{n,\alpha} - 
i m \vec{v}_{F,\alpha} \cdot \vec{v}_s}{\sqrt{\left( 
\tilde{\omega}_{n,\alpha} - i m \vec{v}_{F,\alpha} 
\cdot \vec{v}_s \right)^2 + \tilde{\Delta}_{\alpha}^2}}
\right\rangle_{FS,\alpha}
\end{eqnarray}
where the renormalized Matsubara frequencies $\tilde{\omega}_{n,\alpha}$ and
gaps $\tilde{\Delta}_{\alpha}$ are given by
\begin{equation}
\tilde{\omega}_{n,\alpha} = \Gamma_\alpha \frac{\omega_n}{\sqrt{\omega_n^2+
\Delta_\alpha^2}} \quad {\mathrm{and}} \quad
\tilde{\Delta}_{\alpha} = \Gamma_\alpha \frac{\Delta_\alpha}{\sqrt{\omega_n^2+
\Delta_\alpha^2}}
\end{equation}
Expanding this expression up to third order in $\vec{v}_s$ the sums over
Matsubara frequencies can be done analytically and we find
\begin{eqnarray}
b_\alpha(T) &=&  g_\alpha \frac{3 \pi}{128} \frac{n_\alpha^3}{n_{s0}^3(T)} 
\frac{\Delta_\alpha^2(0) \Delta_\alpha(T) }{\Gamma_\alpha^3} \\
&& \left( 5 \frac{\Delta_\alpha(T)}{T} {\mathrm{sech}}^2 \frac{\Delta_\alpha(T)}{2T} +
6 \tanh \frac{\Delta_\alpha(T)}{2T} \right) \nonumber
\end{eqnarray}
with
\begin{equation}
 n_{s0}(T) = \sum_{\alpha=\pi,\sigma} n_\alpha 
\frac{\pi \Delta_\alpha(T) }{4 \Gamma_\alpha} \tanh \frac{\Delta_\alpha(T)}{2T}
\end{equation}
In the dirty limit the total nonlinear response coefficient $b(T)=b_\pi(T) + 
\frac{j_{c,\pi}^2}{j_{c,\sigma}^2} b_\sigma(T)$ only depends
on the relative ratio of the two scattering rates $\Gamma_\pi/\Gamma_\sigma$.
In Fig.~\ref{fig3} we show its temperature dependence for 
$\Gamma_\pi/\Gamma_\sigma=$0.7, 1, and 2 along with the clean limit result.
As can be seen, the nonlinear response depends on the relative
scattering rates in the two bands. In certain cases the coefficient $b$
can even become smaller than in the clean limit. 

\begin{figure}
  \begin{center}
    \includegraphics[width=0.65\columnwidth,angle=270]{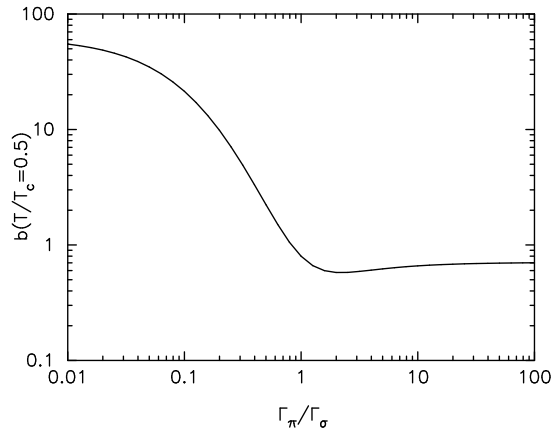}
    \caption{Dependence of $b$ on ratio $\Gamma_\pi/\Gamma_\sigma$ at 
	a fixed temperature of $T=0.5T_c$.
	For $\Gamma_\pi \gg \Gamma_\sigma$ the nonlinear response
	is dominated by the $\sigma$ band, while for 
	$\Gamma_\pi \ll \Gamma_\sigma$ the $\pi$ band dominates.
     \label{fig4} }
  \end{center}
\end{figure}

In order to have more 
insight into this unexpected behavior, in Fig.~\ref{fig4} we plot $b$
at a fixed temperature,  $T=0.5 T_c$, as a function of the
ratio $\Gamma_\pi/\Gamma_\sigma$ on a double logarithmic scale.
We find that $b$ can vary by a factor of order 100 at this temperature.
Qualitatively we can understand this behavior as follows: the
current density is dominated by the band with the smaller
scattering rate, because this band provides the highest conducting
channel. If the $\pi$ band scattering rate is smaller, the total
response is dominated by the small gap leading to a larger nonlinear
response at finite temperature. If, however, the $\sigma$ band 
scattering rate is smaller,the nonlinear response is dominated by the 
large gap giving a smaller nonlinear response.

This analysis tells us that optimization of material properties
in MgB$_2$ should aim at a higher scattering rate in the $\pi$ band
in order to suppress the contribution of the small gap relative
to that of the large gap in the $\sigma$ band. This could be 
achieved by substitutional
doping at the Mg site, for example with Aluminum \cite{Mazin}.
Our analysis also shows that clean MgB$_2$ does not
necessarily provide the lowest nonlinear response.

As has been shown in \cite{Dahm2} the intermodulation power
emitted by a microstrip resonator also depends on the penetration 
depth of the material. A shorter penetration depth leads to an 
increased current density at the edges of the resonator, which
increases the intermodulation power. According to Refs.~\cite{Dahm2}
and \cite{Dahm3} the
intermodulation power $P_{\mathrm{IMD}}$ scales like
\begin{equation}
P_{\mathrm{IMD}} \propto \left( \Delta {\cal L} \right)^2 b^2(T)
\label{pimd}
\end{equation}
where $\Delta {\cal L}$ is a nonlinear coefficient depending on both
the geometry of the resonator and the penetration depth.
According to band structure calculations, the clean limit zero
temperature (London) penetration depth in MgB$_2$ is expected to be near 
$\lambda_L(0)=$40~nm \cite{Golubov}. Actual values vary between
60 and 200~nm depending on film quality \cite{JinAPL,Golubov}.
Comparing MgB$_2$ with the intrinsic $d$-wave
response of YBCO at a temperature of 20~K, we take
$\lambda(T=20K) \approx$ 100~nm for MgB$_2$ and 
$\lambda(T=20K) \approx$ 160~nm for YBCO. Assuming a typical film
thickness of $t=400$~nm we find for the microstrip geometry considered
in Ref.~\cite{Dahm2} $\Delta {\cal L}(t/\lambda=4)=0.173$ (MgB$_2$)
and $\Delta {\cal L}(t/\lambda=2.5)=0.124$ (YBCO). Taking $b(T=20K)=0.563$
for MgB$_2$ and $b(T=20K)=0.855$ for YBCO from Fig.~\ref{fig2} and 
using Eq.~\eqref{pimd}
this means that the intermodulation power in YBCO at this temperature
would be larger by only a factor of 1.2, i.e. a comparable
nonlinear response for both materials.
However, due to their ceramic nature, weak links play a much larger
role in the high-$T_c$ cuprates than in MgB$_2$. For this reason we
expect that the intrinsic nonlinear response will be much easier to
achieve in MgB$_2$ films compared with cuprate films.

We would like to thank R.~B.~Hammond, M.~I.~Salkola, and B.~A.~Willemsen 
for valuable discussions. T.~D. acknowledges the hospitality of
the UCSB Physics Department.

\end{document}